\documentclass[12pt,notitlepage,a4paper]{article}
\usepackage{amssymb}
\usepackage{delarray,amsmath,bbm,epsfig}
\usepackage{cite}

\usepackage{color, graphicx} 

\newcommand\fverb{\setbox\pippobox=\hbox\bgroup\verb}
\newcommand\fverbdo{\egroup\medskip\noindent%
                              \fbox{\unhbox\pippobox}\ }
\newcommand\fverbit{\egroup\item[\fbox{\unhbox\pippobox}]}
\newbox\pippobox
\newcommand{\nn}{\nonumber}
\newcommand{\beq} {\begin{equation}}
\newcommand{\eeq} {\end{equation}}
\newcommand{\beqa} {\begin{eqnarray}}
\newcommand{\eeqa} {\end{eqnarray}}

\newcommand{\vek}{{\vec{k}}}

\newcommand{\ie}{{\it i.e.}}

\newcommand{\order}[1]{${\cal O}\left(#1 \right)$}
\newcommand{\morder}[1]{{\cal O}\left(#1 \right)}
\newcommand{\eq}[1]{(\ref{eq:#1})}

\newcommand{\inv}[1]{\frac{1}{#1}}

\newcommand{\la}{\lambda}

\newcommand{\im}{{\rm Im}}
\newcommand{\re}{{\rm Re}}

\newcommand{\mE}{\mathcal{E}}

\newcommand{\half}{\frac{1}{2}}

\newcommand{\be}{\begin{equation}}
\newcommand{\ee}{\end{equation}}
\newcommand{\bea}{\begin{eqnarray}}
\newcommand{\eea}{\end{eqnarray}}

\def\la{\lambda}

\def\om{\omega}

\newcommand{\vev}[1]{\left\langle#1\right\rangle}

\begin{document}
\begin{flushright}
{\it CP$^\textrm{\it 3}$-Origins-2010-35 } \\
HIP-2010-15/TH
\end{flushright}
\vskip 2cm \centerline{\Large {\bf High-energy asymptotics of D-brane decay amplitudes}} \vskip 3mm
\centerline{\Large {\bf from Coulomb gas electrostatics}} \vskip 1cm
\renewcommand{\thefootnote}{\fnsymbol{footnote}}
\centerline{{\bf Niko
Jokela,$^{1,2}$\footnote{najokela@physics.technion.ac.il} Matti
J\"arvinen,$^{3}$\footnote{mjarvine@cp3.sdu.dk} and Esko
Keski-Vakkuri$^{4,5}$\footnote{esko.keski-vakkuri@helsinki.fi}}}
\vskip .5cm \centerline{\it ${}^{1}$ Department of Physics}
\centerline{\it Technion, Haifa 32000, Israel}
\centerline{\it
${}^{2}$ Department of Mathematics and Physics}
\centerline{\it University of Haifa at Oranim, Tivon 36006, Israel}
\centerline{\it ${}^{3}$CP$^3$-Origins, 
}
\centerline{\it Campusvej 55, DK-5230 Odense M, Denmark}
\centerline{\it
${}^{4}$Helsinki Institute of Physics and ${}^{5}$Department of
Physics } \centerline{\it P.O.Box 64, FIN-00014 University of
Helsinki, Finland}

\setcounter{footnote}{0}
\renewcommand{\thefootnote}{\arabic{footnote}}

\begin{abstract}
We study the high-energy limit of tree-level string production amplitudes from decaying D-branes in bosonic string theory, interpreting the vertex operators as external charges interacting with a Coulomb gas corresponding to the rolling tachyon background, and performing an electrostatic analysis. In particular, we consider two open string - one closed string amplitudes and four open string amplitudes, and calculate explicit formulas for the leading exponential behavior.
\end{abstract}

\newpage



\section{Introduction}

A famous feature of string theory is the soft behavior of scattering amplitudes in the high-energy limit. Remarkably, at each order (worldsheet genus) in the perturbation
theory, the dominant saddle point contribution has a simple electrostatic interpretation - the exponent can be identified as the electrostatic energy of point charges at the equilibrium \cite{Gross:1987kza,Gross:1987ar,Gross:1989ge}. Recently, the electrostatic approach was applied to
string scattering from decaying D-branes \cite{Jokela:2009fd,Jokela:2010nn}.
In the original work \cite{Gross:1987kza,Gross:1989ge,Gross:1987ar}, there was no background charge in addition to the point charges, due to momentum conservation.
A new ingredient for strings scattering from a decaying D-brane is the condensing tachyon on the brane, which provides a background charge distribution that
interacts with the point charges. More precisely, the background consists of a Coulomb gas of unit charges at finite temperature and (imaginary) chemical potential.
This electrostatic interpretation is valid at all energies.
Exact calculations are very complicated, but simplifications can be found in the high-energy limit.
In a previous work, we applied this approach to study closed string pair production from a decaying D-brane at disk amplitude level \cite{Jokela:2009fd}.

In this paper, we study further examples of the electrostatic approach to decay amplitude calculations.
We start by a detailed general description of the approach.
For a uniform discussion, we 
review previous calculations before applying the method to new cases.
We verify that the approach predicts the known high energy behavior of the bulk-boundary amplitude \cite{VEPA}, and show that the amplitude does not diverge at high energies in the kinematically allowed region. We also
present an improved analysis of the (``antipodal'') two-point function, related to open string pair production by the decaying brane.
Then we
discuss $n$-point functions with $n>2$, and the associated string production amplitudes.\footnote{In this paper we consider only tachyon vertex operators.}
As a
completely 
new result, we apply our method to calculate the high-energy limit of a three-point amplitude involving a pair of open strings and a closed string.
For stable D-branes,
such amplitudes are, {\em e.g.}, relevant for emission from or absortion by a charged black hole \cite{Das:1996jy}.
Finally, we extend previous work on $n$-point boundary functions \cite{npt} and study the
high-energy asymptotic behavior of the amplitude for production of four open strings, before
ending with conclusions and outlook. Technical details are presented in two appendices.




\section{Amplitudes from electrostatics approach} \label{sec:string}

\subsection{Preliminaries} 

We review quickly the structure of the disk amplitudes for string scattering from a decaying D-brane in bosonic string theory, with the ``half S-brane''
rolling tachyon background \cite{Sen:2002nu,Larsen:2002wc}. The correlation functions are path integrals
\be
  A_n(\{\omega_a,\vec k_a\}_{a=1}^n) = \int\prod_{a=1}^n\frac{d^2 w_a}{2\pi}\vev{\prod_{a=1}^n V(w_a,\bar w_a)e^{-\delta S_{\rm bdry}}}_{\rm free} \ ,
\ee
where the rolling tachyon boundary deformation representing the half S-brane decay mode for the D-brane is
\be\label{eq:deformation}
 \delta S_{\rm bdry} = \la\int dt e^{X^0(t)} .
\ee
The calculational complications arising from rolling tachyon background are similar for all open and closed
 string vertex operators, when they are represented in timelike gauge \cite{Lambert:2003zr}: 
 \be
 V (k)= V_{sp}(\vec{k}) e^{-i\omega X^0+i\vec{k}\cdot \vec{X}} \ ,
 \ee
 where $V_{sp}(\vec{k})$ involves only space directions $\vec{X}$. The contributions from the contractions
 from the spacelike sector are similar to those in scattering from stable (or non-deformed unstable) D-branes, as they do not involve the
 timelike background (\ref{eq:deformation}). Our focus is in developing techniques
 for calculating the contribution from the nontrivial timelike sector. Therefore, for our purposes it is sufficient to focus on scattering amplitudes which involve only
closed and open string tachyon vertex operators
\be
 V(w_a,\bar w_a) = e^{i k_a\cdot X(w_a,\bar w_a)} \ .
\ee
Closed string vertices are placed in the interior of the unit disk $|w_a|<1$, whereas the open string vertices lie at the boundary $w_a = e^{i\tau_a}, \ \tau_a=0\ldots 2\pi$. We will consider open-closed $n$-point amplitudes,  with $n=n_c+n_o$ where $n_c$ ($n_o$) is the number of closed (open) strings. We adopt a notation $\xi_a=-i\omega_a$ and break up the spatial momentum to parallel and perpendicular directions to the unstable D$p$-brane: $\vec k_a = (\vec k_a^{\parallel},\vec k_a^\perp)$.
On-shell conditions for the bosonic closed string tachyons are $k_a^2=\xi_a^2+(\vec k_a)^2 = -4(N_a-1)$ and for the open string tachyons they read $k_a^2 = \xi_a^2+ (\vec k_a^\parallel)^2 = -(N_a-1)$ \cite{VEPA}.
The overall spatial parallel momentum will be conserved: $\sum_{a=1}^n\vec k_a^\parallel=0$.

The worldsheet correlation functions can be evaluated by first isolating the zero modes from the oscillators, $X^\mu=x^\mu + X'^\mu$, and then expanding the boundary deformation into a power series in $\lambda$. This yields
\bea\label{eq:An}
  A_{n_c+n_o} &=& \int dx^0 d^p\vec x^\parallel e^{i\sum_a k_{a}^\mu x_\mu}\sum_{N=0}^\infty\frac{(-z)^N}{N!}\int\prod_{a=1}^{n_c}\frac{d^2 w_a}{2\pi}\prod_{a=n_c+1}^{n} \frac{d\tau_a}{2\pi} \nn\\
&&\times \prod_{i=1}^N\frac{dt_i}{2\pi}\vev{\prod_{i=1}^N e^{X'^0(t_i)}\prod_{a=1}^n e^{i k_a\cdot X'(w_a,\bar w_a)}} \ ,
\eea
where we introduced $z=2\pi\lambda e^{x^0}$ and fixed the indexing of the vertex operators such that closed strings have smaller values of $a$.

After a straightforward calculation (see Appendix~\ref{app:corre}), we can write (\ref{eq:An}) as
\bea\label{eq:Anno}
  A_{n_c+n_o} & = & \int dx^0 d^p\vec x^\parallel e^{i\sum_a k_{a}^\mu x_\mu}\bar  A_{n_c+n_o}(x^0) \\
\bar A_{n_c+n_o}(x^0) &=& \sum_{N=0}^\infty (-z)^N I_{n_c+n_o}(\{w_a,k_a\};N) \\
& = & \sum_{N=0}^\infty (-z)^N \int\prod_{a=1}^{n_c}\frac{d^2 w_a}{2\pi}\prod_{a=n_c+1}^{n} \frac{d\tau_a}{2\pi}\prod_{1\le a<b \le n} |w_a-w_b|^{2\vec k^\parallel_a\cdot \vec k^\parallel_b} \nonumber\\
&&\times\prod_{1\le a<b\le n_c}|w_a-w_b|^{-k^\parallel_a\cdot k^\parallel_b+\vec k^\perp_a\cdot \vec k^\perp_b}\prod_{a, b=1}^{n_c}|1-w_a \bar w_b|^{\half (k^\parallel_a\cdot k^\parallel_b-\vec k^\perp_a\cdot \vec k^\perp_b)} \nonumber\\ &&\times Z_{n_c+n_o}(\{w_a,k_a\};N) \ , \label{eq:AZrel}
\eea
where
\bea \label{eq:Znndef}
 Z_{n_c+n_o}(\{w_a,k_a\};N) & = & \frac{1}{N!}\prod_{1\le a<b\le n}|w_a-w_b|^{2\xi_a\xi_b} \\
&&\times \int\prod_{i=1}^N\frac{dt_i}{2\pi}\prod_{1\leq i<j\leq N}|e^{it_i}-e^{it_j}|^2\prod_{a=1}^n|1-w_a e^{-it_i}|^{2\xi_a} \ . \nonumber
\eea

\subsection{Applying the electrostatic approach} 

In general, \eq{Znndef} is too complicated to calculate, so we look for an approximation scheme in a special limit. As suggested by the notation, $Z_{n_c+n_o}$ can be interpreted as a Coulomb gas partition function\footnote{We have explored many other aspects of this connection for decaying D-branes in bosonic and superstring
theories in \cite{Balasubramanian:2006sg,Hutasoit:2007wj,Jokela:2007dq,Jokela:2007wi,Jokela:2009gc}.} at the inverse
temperature $\beta=2$. We can evaluate the partition function in a saddle point approximation corresponding to electrostatic equilibrium, at large $N$ and $\xi_a$ (with the ratios $\xi_a/N$ fixed). The method is discussed in detail in a companion paper \cite{Jokela:2010nn}.
The leading term of $\log Z_{n_c+n_o}$ can be found by evaluating the electrostatic energy of the Coulomb gas, unless $n_c+n_o$ is too large. The next-to-leading term was also analyzed in \cite{Jokela:2010nn}.
In this article we shall continue the analysis of \cite{Jokela:2010nn} by discussing in detail how the high energy limit of string amplitudes arises from the electrostatic approach.
String production at large
$\xi_a=-i\omega_a$ 
is a physically interesting regime,
since we expect the unstable D-brane to mainly decay to very massive string modes \cite{Lambert:2003zr}.

First we note that the correlator $\sim Z_{n_c+n_o}$
needs to be summed over $N$ and integrated over $x^0$ to obtain the string scattering amplitude,
\be
 A_{n_c+n_o} \propto \int dx^0 e^{\sum_a\xi_a x^0} \sum_{N=0}^\infty (-z)^N Z_{n_c+n_o}(\{\xi_a\};N) \ ,
\ee
where $z=2\pi\la e^{x^0}$ and we dropped the complicated proportionality factor for clarity. As discussed in \cite{npt}, under certain
conditions the sum and the integral can be carried out exactly,
\be \label{eq:Ampres}
  \int\!\! dx^0 e^{\sum_a\xi_a x^0} \! \sum_{N=0}^\infty (-z)^N Z_{n_c+n_o}(\{\xi_a\};N) =\frac{\pi(2\pi\la)^{-\sum_a \xi_a}}{\sin\left( \pi \sum_a\xi_a\right)} Z_{n_c+n_o}\!\!\left(\!\{\xi_a\};N = -\!\sum_a \xi_a\!\right) \ .
\ee
This result was originally found in \cite{Zamolodchikov:1995aa} in the
context of Liouville field theory.
$Z_{n_c+n_o}$ is in general unknown, but simplifies in the large $N$ limit
($N,\xi_a \to \infty$ with $N/\xi_a$ fixed) \cite{Jokela:2010nn}. Since
$N=-\sum_a\xi_a$ in \eq{Ampres}, the large $N$ limit corresponds to the high
energy ($\om_a = i\xi_a$) limit for the string production amplitudes. In
particular, the leading term gives the saddle-point approximation
\be \label{eq:Ampres2}
 A_{n_c+n_o} = 
 \frac{\pi(2\pi\la)^{-\sum_a \xi_a}}{\sin\left( \pi \sum_a\xi_a\right)} \int (dwd\bar{w} d\tau )\gamma (w,\bar{w},\tau) \exp\left[-2\left.\mE\right|_{ N = -\sum_a \xi_a}\right] \ 
 \ ,
\ee
where $\mE$ is the electrostatic energy of the corresponding Coulomb gas configuration in the continuum limit, and
can be computed explicitly \cite{Jokela:2010nn}.  Above $\int (dwd\bar{w}d\tau)$ denotes the $w_a$
and $\tau_a$ integrals, and $\gamma (w,\bar{w},\tau )$ denotes the contribution from spacelike
contractions in \eq{Anno}.
Their explicit form depends on the particular amplitude. Note that in general also $\mE$ is
$w_a,\tau_a$ dependent. If possible, we
perform the integrals explicitly, but in general we follow the approach of \cite{Gross:1987kza} and replace $w_a$ by their electrostatic equilibrium values
 (for details, see Appendix~\ref{app:equil} and the examples in Sections~\ref{sec:2pt} and~\ref{sec:npt}).

 The use of \eq{Ampres} requires an analytic continuation of $Z_{n_c+n_o}(N)$ in the parameter $N$
 to noninteger $N$, which is subject to some constraints.
 The partition function should be analytic for $\re \, N>0$, and it should not grow exponentially as $N \to \infty $, when $|\arg N|<\pi/2$.
The asymptotic behavior of $Z_{n_c+n_o}$ 
(with $\xi_a$ fixed) may be analyzed using random matrix theory techniques.
Using a random matrix theory interpretation, $Z_{n_c+n_o}$ describes an
expectation value of a periodic function in the circular ensemble of $U(N)$ random matrices, ${\rm{CUE}}(N)$.
The expectation value can in turn be converted to a Toeplitz determinant of the Fourier coefficients of periodic funtion (see \cite{VEPA,VEPA2,Jokela:2007yc} for more discussion), the advantage then is that
the asymptotic behavior of the determinant at $N\rightarrow \infty$ simplifies, and is described by the Szeg\"o formula and its generalization.
Using the result of \cite{Fisher}, we find
the power-law behavior
\be \label{eq:ZNasymp}
 Z_{n_c+n_o}(N)  \substack{\phantom{N \om N} \\ \sim \\ N \to \infty} N^{\sum_{a=n_c+1}^n \xi_a^2} 
\ee
(see \cite{Jokela:2009fd} for a discussion with $n_c=0$).
This result holds also for the leading term in the $N \to \infty$ limit
when $N/\xi_a$ is fixed,
an explicit discussion for the boundary one-point partition function is given in \cite{Jokela:2010nn}.

The above analysis needs to be modified if the large $N$ limit of the partition function is not analytic in $N$.
As explained in \cite{Jokela:2009fd,Jokela:2010nn}, nonanalytic behavior may be associated with suddenly appearing or disappearing gaps\footnote{A gap may also exist first and then disappear, or many gaps may be created
 or join together.} in the continuous charge distribution
of the Coulomb gas picture, 
created in the vicinity of
the external charges at some large value of $N$.
This may happen in the presence of bulk charges: a charge near the boundary of the disk generates a
gap, which disappears at sufficiently large $N$ \cite{Jokela:2009fd}. Similar phenomenon may occur in higher order boundary
amplitudes: two nearby charges will create only one gap for small $N$, which separates into two gaps as $N$ increases. In these
cases, as suggested in \cite{Jokela:2009fd}, one can first integrate over the locations of the charges. The integrated partition
function is analytic in $N$, and one can proceed with the analytic continuation.\footnote{Another method which typically avoids non-analyticities is to fix the positions by the equilibrium equations, which follow from minimizing the electrostatic energy and will be discussed in Appendix B.}

In order to calculate the electrostatic energy, we need to first assume that $\xi_a$ are real valued, and then continue analytically to physical energies $\om_a = i\xi_a$ in the end. We justify this by noting that $Z_{n_c+n_o}$ is analytic for $\re \xi_a \ge -1/2$, as seen from \eq{Znndef}.\footnote{There may be
one caveat: the amplitude contains an integration over the unfixed moduli parameters, which
 might in principle cause problems
 in the case of higher-point amplitudes, 
 since the integrals are not necessarily well defined for all $\re\xi_a \ge 0$. This needs to be checked case by
 case.}

In summary, we have developed a prescription for calculating the high energy approximation of
the string scattering amplitudes, containing the following steps:
\begin{enumerate}
 \item Find the 
 series expansion
 for the scattering amplitude (see Eqs.~(\ref{eq:Anno})-(\ref{eq:Znndef})) and identify the Coulomb gas partition function.
 \item Solve the electrostatic potential problem to find $\mE$, and the leading partition function $\log Z \simeq -2 \mE$ at $\beta=2$.
 \item If necessary, integrate the leading result over the positions of the external charges (the unfixed modular parameters of the string vertex operators).
 \item Do the summation over $N$ and integration over time by continuing analytically $N$ to the total external charge $N \to -\sum_a\xi_a$ in the result for $\log Z$, as shown in Eq.~\eq{Ampres}.
 \item Continue analytically to physical energies $\om_a= i\xi_a$.
\end{enumerate}

Finally, let us comment on the precision of the obtained approximation. We
only used the leading term of $\log Z_{n_c+n_o}$ in the large $N$ limit,
which was shown to be \order{N^2} in \cite{Jokela:2010nn}. Since we fixed
$N= - \sum_a\xi_a$, our result \eq{Ampres2} is the leading \order{\om^2}
term of $\log A_{n_c+n_o}$ in the limit of large $\om = \sum \om_a$ with the
ratios $\om_a/\om_b$ fixed. We also calculated the next-to-leading \order{N}
term of $\log Z_{n_c+n_o}$ in  \cite{Jokela:2010nn}. However, this term
vanishes at $\beta=2$. Therefore, the first nontrivial corrections to our
high-energy approximation are obtained from the \order{N^0} term of $\log
Z_{n_c+n_o}$, resulting in \order{\om^0} (possibly logarithmic) corrections
to $\log A_{n_c+n_o}$.

\subsection{Electrostatic energies}

In the remainder of the paper we study the easiest string scattering amplitudes following the above
systematic prescription.
In the second step, we need the electrostatic potential energy $\mE$, as in Eq.~\eq{Ampres2}. We have
already studied this problem in \cite{Jokela:2009fd,Jokela:2010nn}, and summarize the results here.
First, the energy with one bulk charge $\xi=\xi_1$ at $w=w_1=r$  ($0<r<1$) reads \cite{Jokela:2009fd}
\bea \label{eq:bulk1}
 \mE^{({\rm bulk})} &=& \theta(r_c-r) \frac{\xi^2}{2} \log\left(1-r^2\right) \\
& &+ \theta(r-r_c) \left[  - \frac{(N+2 \xi)^2}{4}\log
\frac{1+\chi}{1+\delta(r)}-\frac{N^2}{4}\log
\frac{1-\chi}{1-\delta(r)} \right. \nonumber \\
& & \left. \ \ \ \ \ \ \ \ \ \ \ \ \ +\frac{\xi^2}{2}\log\frac{4\chi}{(1+\delta(r))^2}
 \right] \nonumber \ ,
\eea
where
\be\label{eq:rc}
r_c=\frac{N}{N+2\xi} \ ; \
\delta(r) = \frac{1-r}{1+r} \ ; \ \chi= \frac{\xi}{N+\xi} \ .
\ee
Several configurations with charges on the boundary of the disk were solved in \cite{Jokela:2010nn}. In the simplest case there is one boundary charge $\xi$ on the boundary, giving
\bea \label{eq:bdry1}
 \mE^{({\rm 1pt})} &=& \inv{2}\left[-2F(\xi)+ F(2\xi) - F(N) - F(N+2\xi)+ 2F(N+\xi)\right] \ ,
\eea
where
\be
 F(x) = \inv{2}x f(x) = \inv{2} x^2 \log x \ .
\ee
We will also use the result for the configuration where two boundary charges $\xi_1$ and $\xi_2$ lie at exactly antipodal points on the unit circle. Then
\bea \label{eq:bdry2}
 \mE^{({\rm 2pt})} &=& \bigg\{F(N+\xi_1+\xi_2)- F(\xi_1+\xi_2) \nn\\
  &  & - \inv{4}\left[F(N+2 \xi_1+2\xi_2) +F(N+2 \xi_1) +F(N+2 \xi_2) +F(N)\right] \nn \\
       &   & + \inv{4}\left[ F(2 \xi_1+ 2 \xi_2)+F(2 \xi_1)+ F(2 \xi_2)\right] - \xi_1\xi_2\log 2 \bigg\} \ .
\eea
As an example of a more complicated situation we consider a symmetric four-point case where two particles with
equal charges $\xi_1$  are, say, at $\tau=\arg w = 0$ and at $\tau=\pi$, and two additional particles with
charges $\xi_2$ are at $\tau=\pi/2$ and at $\tau=3\pi/2$. This configuration gives
\bea  \label{eq:bdry4}
 \mE^{({\rm 4pt})} 
 & = & 2 F(N/2+\xi_1+\xi_2) - 2 F(\xi_1+\xi_2) \nn\\
  &  & - \inv{2}\left[F(N/2+2 \xi_1+2\xi_2) +F(N/2+2 \xi_1) +F(N/2+2 \xi_2) +F(N/2)\right] \nn \\
       &   & + \inv{2}\left[ F(2 \xi_1+ 2 \xi_2)+F(2 \xi_1)+ F(2 \xi_2)\right] -\left(\xi_1+\xi_2\right)^2\log 2 \ .
\eea


\section{Two-point amplitudes} \label{sec:2pt}

\subsection{Bulk-bulk amplitude}
The high-energy production amplitude of two closed strings from a decaying D-brane
was derived in \cite{Jokela:2009fd} by using 
the framework of Section~\ref{sec:string} with the result of Eq.~\eq{bulk1}.
For completeness, we review the results here. 
We fix the charge $\xi_2$ at $w_2=0$, and the charge $\xi_1$ at $0<w_1=r<1$ by using conformal symmetry. In this case Eq.~\eq{Ampres2} becomes
\be
 A_{0+2}(\xi_1,\xi_2) = \frac{\pi(2\pi\la)^{-\xi_1-\xi_2}}{\sin\pi(\xi_1+\xi_2)} \left\{\int_0^1 dr k(r) \exp\left[-2\mE^{\rm (bulk)}\right]\right\}_{N=-\xi_1-\xi_2} \ ,
\ee
where we restored the proportionality factor
\be \label{eq:kdef}
 k(r) = r r^{k_1\cdot k_2}(1-r^2)^{\inv{2}\left((k_1^{||})^2-(\vek_1^{\perp})^2 \right)} \ .
\ee
For the special case where the strings have equal energies, $\om_1=\om_2\equiv \om$, the result simplifies to
\be \label{eq:bulkamp}
 A_{0+2}(\om) = -\frac{ i \pi e^{i\frac{\pi s}{2}}(2\pi\la)^{2 i\omega}}{2\sinh2\pi\omega}\left[e^{i\frac{\pi u}{2}}\frac{\Gamma\left(\frac{t}{2}-1\right)}{\Gamma\left(\frac{t+u}{2}-2\right)}+\frac{\Gamma\left(\frac{s}{2}+1\right)}{\Gamma\left(\frac{s+u}{2}\right)}\right]\Gamma\left(\frac{u}{2}-1\right) \ ,
\ee
where
\bea
 s &=& k_1\cdot k_2 \nn\\
 t &=& 2 (k_1^{||})^2 = 2 (k_2^{||})^2 = 8 - (\vek_1^{\perp})^2-(\vek_2^{\perp})^2 \nn\\
 u &=& \frac{(\vek_1^{\perp}-\vek_2^{\perp})^2}{2} \ .
\eea
In the definitions of the parameters we used the on-shell conditions $k_a^2=-\om_a^2+\vek_a^2=-4(N_a-1)$  for the tachyonic states $N_a=0$. Notice that the results for the leading high energy asymptotics (see \cite{Jokela:2009fd} for an extensive analysis) remain valid also for higher $N_a$'s as long as they are much smaller than the (squared) energy scale.

\subsection{Bulk-boundary amplitude}

Let us then discuss the asymptotics of the bulk-boundary scattering amplitude,
and verify that our result mathces the exact one \cite{VEPA}. 
We follow \cite{Lambert:2003zr,VEPA,VEPA2} and use conformal symmetry to place the bulk operator (charge $\xi_c$) to the origin, rather than
integrating over its position as suggested in \eq{AZrel}. Then the bulk charge decouples from the Coulomb gas calculation, and we may use the
results with one boundary charge ($\xi_o$) for the partition function. According to Eq. \eq{Ampres2}, analytic continuation of \eq{bdry1}
to $N=-\xi_c - \xi_o$  at $\beta=2$ gives
\be \label{eq:1ptAas}
   A_{1+1}(\xi_c,\xi_o) \simeq  \frac{\pi(2\pi\la)^{-\xi_c-\xi_o}}{\sin\pi\left(\xi_c+\xi_o\right)} \exp[-2\left.\mE^{({\rm 1pt})}]\right|_{\xi=\xi_o,\, N = -\xi_c-\xi_o} \ .
 \ee
Setting here $\xi_a = -i\om_a$ we find the asymptotics
\bea \label{eq:11as}
   A_{1+1}(\om_o,\om_c)  &=& \frac{i\pi\left(2\pi\la\right)^{i\left(\om_c+\om_o\right)}}{\sinh \pi \left(\om_c+\om_o\right)} \exp\left[ \om_o^2 \log\frac{\om_o}{\om_c+\om_o} + \om_c^2 \log\frac{\om_c}{\om_c+\om_o}\right. \nonumber \\
&& - \left.\inv{2}\left(\om_c-\om_o\right)^2\log\frac{\om_c-\om_o}{\om_c+\om_o} +\om_o^2\left(2\log 2\pm i\pi\right) + \morder{\log\om}\right] \ ,
\eea
where we restored the expected size of the next-to-leading order correction.
This indeed matches with the asymptotics of the exact amplitude \cite{VEPA} up to the branch choice of the
logarithm ($\pm$ in the phase factor, on the last line in \eq{11as}) which is hard to obtain from the electrostatic approach. Notice, however, that the
absolute value of the amplitude is independent of the branch.

Let us make one comment about this result. After using momentum conservation parallel to the D-brane, 26-momenta of the strings become
\bea
 k_c &=& (\om_c,\vek^\parallel,\vek^\perp) \nn\\
 k_o &=& (\om_o,-\vek^\parallel,0) \ .
\eea
At high energy,
and for low-lying excitations ($N_a \ll \om_a^2$), the mass-shell conditions $-\om_c^2 + \vek_c^2 =-4(N_c-1)$ and $-\om_o^2 + \vek_o^2 =-(N_o-1)$ 
give
\bea
 \om_c &\simeq& \sqrt{\left(\vek^\parallel\right)^2+\left(\vek^\perp\right)^2} \nn\\
 \om_o &\simeq& \left|\vek^\parallel\right| \ ,
\eea
so asymptotically $\om_c\ge \om_o$. The leading term in \eq{11as} can be written as
\bea \label{eq:11lead}
 &&\om_o^2 \log\frac{\om_o}{\om_c+\om_o} + \om_c^2 \log\frac{\om_c}{\om_c+\om_o}
 - \inv{2}\left(\om_c-\om_o\right)^2\log\frac{\om_c-\om_o}{\om_c+\om_o} +2 \om_o^2\log 2 \\\nn
 &=& (\om_c +\om_o)^2\left[\alpha^2 \log \alpha + (1\!-\!\alpha)^2 \log(1\!-\! \alpha) -\inv{2}\left(1\!-\!2\alpha\right)^2\log(1\!-\!2\alpha) +2\alpha^2\log 2 \right] \ ,
\eea
where $\alpha \equiv \om_o/(\om_c+\om_o)$. In the kinematically allowed region $0<\alpha< 1/2$ the function in the
square brackets in \eq{11lead} is negative, and it vanishes at the endpoints $\alpha=0,1/2$. Thus the amplitude vanishes for
high energies in the kinematically allowed region as $A\sim e^{- \pi(\om_c+\om_o)}$ if $\om_c \gg \om_o$
or $\om_c=\om_o$, 
and faster  ($\sim e^{-\#(\om_c+\om_o)^2}$) if the energies are comparable
but inequal. 
We observed similar behavior for the bulk
two-point amplitude \eq{bulkamp} at high energies in \cite{Jokela:2009fd}.

\subsection{Boundary-boundary amplitude}

Finally, we shall analyze the boundary two-point amplitude.
Momentum conservation fixes $\vek_1^\parallel=-\vek_2^\parallel\equiv \vek^\parallel$
and from the on-shell conditions for low mass excitations $-\om_a^2 + \vek_a^2 = \morder{\om^0}$  we get
$\om_1\simeq\om_2\equiv\om$ at high energy.
The electrostatic two-point partition function at equal charges was found only numerically for general $\tau=\tau_2-\tau_1$ in \cite{Jokela:2010nn}. Therefore, we shall calculate the amplitude in the equilibrium configuration. Notice that due to symmetry the
configuration where the charges lie at antipodal points, $\tau_2=\tau_1+\pi$, is always a solution to the equilibrium
equations \eq{eqeq}: if we set $\tau_1=0$ and $\tau_2=\pi$ the charge distribution is symmetric
with respect to the real axis, and the imaginary parts in both of the terms of \eq{eqeq} vanish.
The total energy is found by setting $\xi_1=\xi_2$ in Eq.~\eq{bdry2}
which yields
\be
 \mE^{\rm(2pt, s)} = \inv{2} F(N+2\xi) - \inv{4} F(N) - \inv{4} F(N+4\xi) + \xi^2\log 4\xi \ .
\ee
Including the spatial momentum dependence from Eq.~\eq{AZrel}, and by using Eq.~\eq{Ampres},
we find
\bea
 A_{0+2}(\om) \simeq \frac{i\pi\left(2\pi\la\right)^{2i\om}}{\sinh 2\pi\om} 2^{-2\left(\vek^\parallel\right)^2} \exp\left(-2\left.\mE^{\rm(2pt, s)}\right|_{N=2i\om,\xi=-i\om}\right)
\eea
After using the on-shell condition $-\om^2+\vek^2_\parallel=\morder{\om^0}$ the amplitude becomes
\be\label{eq:boundbound}
  {A}_{0+2}(\om) = \frac{i\pi\left(2\pi\la\right)^{2 i \om}}{\sinh 2 \pi \om} e^{\pm i\pi\om^2 + \morder{\log\om}} \ .
\ee
The result vanishes for large energies as $A\sim e^{-2\pi\om}$. It is in accord with the one suggested in \cite{Gutperle:2003xf} and also matches with
the bulk-boundary amplitude \eq{11as} asymptotically at $\om_c=\om_o$.


\section{Higher-point amplitudes} \label{sec:npt}

\subsection{Bulk-boundary-boundary amplitude}

We can extend our method also to the three point amplitude $A_{1+2}$. As above, we place the bulk charge at the origin, where it decouples from the
Coulomb gas analysis. We then fix the boundary charges at the equilibrium configuration, where they are at antipodal points on the circle. The
relevant total energy is given by Eq.~\eq{bdry2}. By using Eqs.~\eq{AZrel} and \eq{Ampres}, we find
\be
 A_{1+2}(\om_c,\om_1,\om_2) \simeq \frac{i\pi\left(2\pi\la\right)^{ i\sum_a \om_a}}{\sinh\left( \pi\sum_a \om_a\right)} 2^{2\vek_1^\parallel\cdot\vek_2^\parallel} \exp\left[-2\left.\mE^{\rm(2pt)}\right|_{N=i\sum_a\om_a,\ \xi_a=-i\om_a}\right] \ ,
\ee
where the subscripts $1,2$ ($c$) refer to the open strings (closed string), and $a=1,2,c$ in the sums.
The 26-momenta can be written as
\bea
 k_1 &=& (\om_1,\vek_1^\parallel,0) \nn\\
 k_2 &=& (\om_2,\vek_2^\parallel,0) \nn\\
 k_c &=& (\om_c,-\vek_1^\parallel-\vek_2^\parallel,\vek^\perp) \ .
\eea
At high energies mass-shell conditions give $\om_a\simeq \left|\vek_a^\parallel\right|$, $a=1,2$, and therefore
\be \label{eq:ineq}
 \om_c^2 \simeq \left(\vek_1^\parallel+\vek_2^\parallel\right)^2+\left(\vek^\perp\right)^2 \ge \om_1^2 + \om_2^2 + 2 \om_1 \om_2 \cos \phi  \ge \left( \om_1- \om_2\right)^2 \ ,
\ee
where $\phi$ is the angle between the vectors $\vek_1^\parallel$ and $\vek_2^\parallel$.
The result for the amplitude may be written as
\bea
 A_{1+2}(\om_c,\om_1,\om_2)  &=& \frac{i\pi\left(2\pi\la\right)^{ i\sum_a \om_a}}{\sinh \left(\pi\sum_a \om_a\right)}\\ \nn
 &&\times \exp\!\left\{\left(\om_1\!+\!\om_2\!+\!\om_c \right)^2\!\left[\Xi(\alpha_1,\alpha_2,\phi) \pm i\pi \Theta(\alpha_1,\alpha_2) \right] + \morder{\log \om} \right\} \ .
\eea
Here the functions $\Xi$ and $\Theta$ are defined as
\bea
 \Xi(\alpha_1,\alpha_2,\phi) &=&  - \inv{2}\left[F(|1\!-\!2 \alpha_1\!-\!2\alpha_2|) +F(1\!-\!2 \alpha_1) +F(1\!-\!2 \alpha_2)\right] \nn \\
       &   & + \inv{2}\left[ F(2 \alpha_1\!+\! 2 \alpha_2)+F(2 \alpha_1)+ F(2 \alpha_2)\right] \nn\\
       & & - 2 F(\alpha_1\!+\!\alpha_2) + 2 F(1\!-\!\alpha_1\!-\!\alpha_2) -2\left(1\!-\!\cos\phi\right) \alpha_1\alpha_2 \log 2 \\
    \Theta(\alpha_1,\alpha_2) &=& \alpha_1^2+\alpha_2^2 - \inv{4}\left(1\!-\!2\alpha_1\!-\!2\alpha_2\right)^2\theta\left(2\alpha_1\!+\!2\alpha_2\!-\!1\right)  \ ,
\eea
where $F(x) = (x^2\log x)/2$ as usual, $\theta$ is the step function, and $\alpha_a = \om_a/\sum_b\om_b$. Notice that the second inequality in Eq.~\eq{ineq} restricts $0\le\alpha_a\le 1/2$,
\ie, neither of the open strings can alone carry more than half of the total emitted energy.

\begin{figure}[th]
\centering
\includegraphics[width=0.6\textwidth]{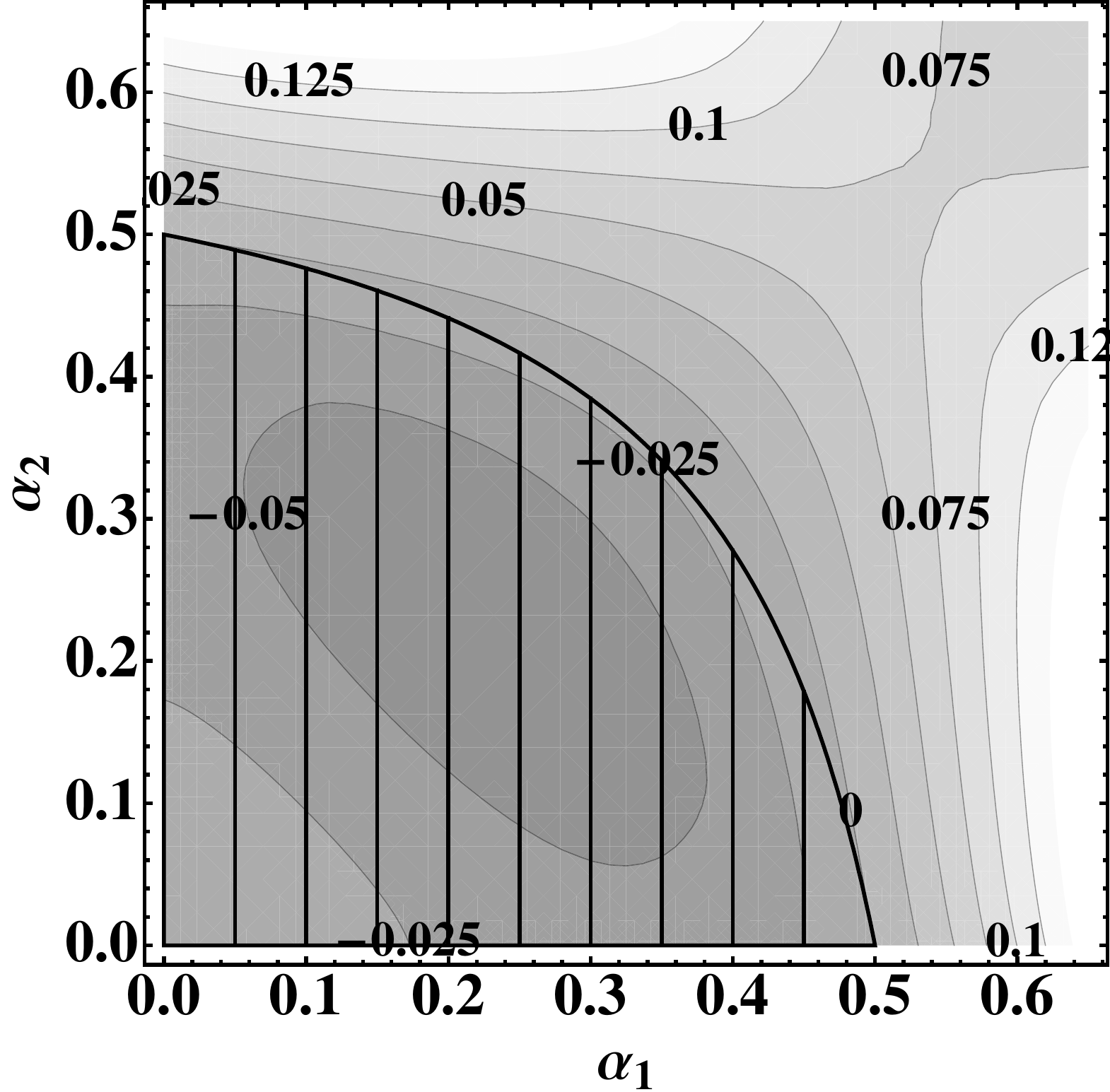}%
\caption{The function $\Xi$ for $\cos\phi=0.6$. The curves lie at constant $\Xi$ as indicated by the labels, and the striped region is the kinematically allowed one, given by Eq.~\eq{ineq2}.}
\label{fig:negativity}
\end{figure}

It is crucial that the function $\Xi$ is negative
in the physical region for the result to make sense: the amplitude must not diverge at high energies. The first inequality in Eq.~\eq{ineq} may be written as
\be \label{eq:ineq2}
 2\alpha_1\alpha_2\left(1\!-\!\cos\phi\right) \ge 2\alpha_1+2\alpha_2-1 \ .
\ee
We verified numerically that $\Xi$ is negative (or zero) whenever this inequality is satisfied.
Fig.~\ref{fig:negativity} shows the situation for $\cos\phi=0.6$: $\Xi$ is indeed negative in the whole kinematically allowed (striped) region.
This is a most nontrivial check of our result.
Notice that
$\Xi$ vanishes if $\cos\phi =\pm 1$ and Eq.~\eq{ineq2} is saturated. In fact, $\Xi$ vanishes if and only if the spatial momenta of all the strings are (asymptotically) parallel, with the understanding that zero is always parallel to any vector. Thus the amplitude decays exponentially
\be
 A_{1+2}(\om_c,\om_1,\om_2) \sim e^{-\pi\sum_a \om_a}
\ee
for large energies in these configurations, and the decay is even faster
\be
 A_{1+2}(\om_c,\om_1,\om_2) \sim \exp\bigg[-\#\Big(\sum_a\om_a\Big)^2\bigg]
\ee
in other cases.

\subsection{Boundary amplitudes}
We can also give other conjectures on the asymptotics of $n$-point amplitudes with $n \ge 3$. We are, however, limited to special kinematic settings,
which can be accessed by solutions  for the symmetric potential problems in \cite{Jokela:2010nn}. For example, let us consider ``pairwise'' production
of four open strings, with $\vek_1^\parallel=-\vek_3^\parallel$, and  $\vek_2^\parallel=-\vek_4^\parallel$, and with possibly different
energies $\om_a \simeq \left|\vek_a^\parallel\right|$, $a=1,2$. We fix the charges symmetrically as explained above before Eq.~\eq{bdry4}.
In the case of production to orthogonal directions ($\vek_1^\parallel \cdot \vek_2^\parallel=0$)
this configuration also solves the equilibrium equations \eq{eqeq}.
Proceeding as above with $\mE^{\rm (4pt)}$ from Eq.~\eq{bdry4},
\bea
 A_{0+4}(\om_1,\om_2) &\simeq& \frac{i\pi\left(2\pi\la\right)^{ 2i( \om_1+\om_2)}}{\sinh\left[2 \pi\left(\om_1+\om_2\right)\right]} \exp\left[F(2\om_1)+F(2\om_2)\right.\nn\\
 &&\left.-2F(\om_1+\om_2)-2F(|\om_1-\om_2|) -2 \left(\om_1^2+\om_2^2\right)\log 2\right] \ .
\eea
One can check that the result vanishes asymptotically for any fixed ratio $\om_1/\om_2$ as $A\sim e^{-\#\left(\om_1+\om_2\right)^2}$.


\section{Conclusions and outlook}\label{sec:discussion}

In this paper we used electrostatic techniques to investigate
string scattering amplitudes of D-brane decay in bosonic string
theory.
In particular, we studied the high energy limits of open and closed
string emission amplitudes in the half S-brane background. We considered
pair production of open strings and closed
strings. We also derived a result for a mixed amplitude with one closed
string
and a pair of open strings, and briefly discussed $n$-point open string amplitudes
with $n\geq 3$.

Overall, our analysis revealed the expected exponential fall-off behavior at high
energies -- the amplitudes decay with sums of the emitted energies in
the exponent. However, in many cases
the decay was found to be even faster, depending on the conditions for the spatial
momenta.

An attractive feature of the electrostatic method is that it provides intuitive
insight into the high-energy behavior of the string amplitudes. It would be worthwhile
to generalize our investigations to other unstable systems.  Some cases
to study are: 1) full S-brane, which corresponds to a
collection of positive and
negative unit charges \cite{Jokela:2007wi}, 2) non-BPS D$p$-branes for $p$ odd/even in
Type IIA/B superstring, corresponding to paired Coulomb gases
\cite{Hutasoit:2007wj},
and 3) inhomogeneous or lightlike decays, possibly corresponding to two sets of distinct Coulomb
gases. The continuum limit with appropriate external
bulk or boundary charges in each case
would help to find an approximate high energy emission
amplitude for closed or open strings.

The high energy closed string pair
production amplitude which we obtained is currently the only explicit result
for this process. There are two remaining
puzzles which we have so far failed to solve. First,
in order to maintain symmetry in exchanging the closed strings
we had to fix the energies of the closed strings
to be equal \cite{Jokela:2009fd}.
This requirement is a limitation. It does not arise from the electrostatic
approximation -- in the exact power series
expression of the
amplitude, each term in the expansion is by itself asymmetric.\footnote{Notice that each term in the expansion is highly off-shell, so one does not really
expect individual terms to be manifestly exchange symmetric.} However, it would be natural for the final amplitude to depend symmetrically on the energies of the emitted closed strings.


The second puzzle is associated with
open-closed duality \cite{Sen:2003xs}. (For discussion on the issue, see {\em e.g.} \cite{Karczmarek:2003xm,Sen:2004nf,Nakayama:2006qm,Song:2010hc}.)
A striking mismatch of the duality is the failure to connect the closed string IR channel to the open string UV channel. Consider
an open string annulus diagram, with rolling tachyon background on a decaying $p$-brane at both edges. The two natural ways
of cutting the annulus, and the two corresponding different kinematical limits, give total amplitudes
for open string or closed string pair production, with UV region of the open string channel
corresponding to the IR region in the closed string channel \cite{Karczmarek:2003xm}. However, the
closed string  production rate is finite for $p\leq 23$, whereas the open string pair production rate in
the UV is finite for $p\leq 22$. In an ongoing work, we have tried to improve the open string analysis by including
logarithmic corrections to the exponent, using the electrostatic approach, as in
(\ref{eq:boundbound}). However, the
mismatch between the open string and closed string production rates seems to become even more
pronounced.  We have estimated the correction numerically and
have found the open string pair production rate
become UV finite for all D$p$-branes, with a bound $p\leq
26$.\footnote{The emitted energy still diverges for $p=25$.} So the question remains,
is it valid to think of the bulk one-point and the boundary two-point
amplitudes as coming from the same vacuum
open string one-loop amplitude?

There are some caveats.
First, there is no rigorous justification of the analytic continuation method from Euclidean to Minkowski signature,
proposed in \cite{Gutperle:2003xf}, to obtain the exact open string
pair production amplitude. Furthermore, it seems to be
very difficult to extract the IR limit of the open string amplitude
in \cite{Gutperle:2003xf}, in order to make
contact with the closed string UV channel. As far as we know, no
results in the open string IR channel are known.
Second, there are no results for amplitudes in superstring theory beyond
the bulk-boundary amplitude \cite{VEPA2}. Bulk one-point amplitudes have been calculated in \cite{Shelton:2004ij}, and the closed string
production rate in the IR can be easily extracted to be finite for
$p\leq 7$. It would be interesting to generalize our electrostatics methods to Type II
superstring and find out how the open string pair production amplitude
behaves at high energies.

\bigskip
\noindent

{\bf \large Acknowledgments}

We wish to thank Oren Bergman and Gilad Lifschytz for many useful discussions. N.J. has been supported in part by the Israel Science
Foundation under grant no. 568/05 and in part at the Technion by a fellowship from the Lady Davis Foundation. M.J. has been supported in part by the
Villum Kann Rasmussen Foundation. E.K-V. has been supported in part by the Academy of Finland grant number
1127482. This work has also been supported in part by the EU 6th Framework Marie Curie Research and Training network ``UniverseNet''
(MRTN-CT-2006-035863).

\appendix

\section{Various contractions}\label{app:corre}

In this appendix we will fill in some gaps between (\ref{eq:An}) and (\ref{eq:Anno}). For ease of reference, let us record (\ref{eq:An}),
\bea\label{eq:Annew}
  A_{n_c+n_o} &=& \int dx^0 d^p\vec x_\parallel e^{i\sum_a k^{(a)}_\mu x^\mu}\sum_{N=0}^\infty\frac{(-z)^N}{N!}\int\prod_{a=1}^{n_c}\frac{d^2 w_a}{2\pi}\prod_{a=n_c+1}^{n} \frac{d\tau_a}{2\pi} \nn\\
&&\times \prod_{i=1}^N\frac{dt_i}{2\pi}\vev{\prod_{i=1}^N e^{X'^0(t_i)}\prod_{a=1}^{n_c+n_o} e^{i k_a\cdot X'(w_a,\bar w_a)}} \ .
\eea

To calculate the full contraction, it is useful to include various contributions one by one. Denote
\begin{equation}
 K_{n_c+n_o} \equiv \vev{\prod_{a=1}^{n_c+n_o} e^{i k_a\cdot X'(w_a,\bar w_a)}} \ .
\end{equation}
Let us first focus just on closed strings, $n_o=0$.  It is important to recall that the closed strings have mixed boundary conditions, Neumann for parallel ones $(\mu=0,\ldots,p)$ and Dirichlet for perpendicular directions $(\mu=p+1,\ldots,25)$. This is encoded
in the Green's functions \cite{VEPA}
\be
 \vev{X'^{\mu}(w_a,\bar w_a)X'^{\nu}(w_b,\bar w_b)} = \left\{ \begin{array}{ll}
-\half\eta^{\mu\nu}(\log|w_a-w_b|^2+\log|1-w_a\bar w_b|^2)\ , & \textrm{\ Neumann} \\
-\half\eta^{\mu\nu}(\log|w_a-w_b|^2-\log|1-w_a\bar w_b|^2)\ , & \textrm{\ Dirichlet\ .}
\end{array} \right. \label{eq:corre}
\ee
These yield (the singular self-contractions are dropped)
\be\label{eq:Knc}
 K_{n_c+0} = \prod_{1\leq a<b\leq n_c}|1-w_a\bar w_b|^{k^\parallel_a\cdot k^\parallel_b-\vec k^\perp_a\cdot\vec k^\perp_b}|w_a-w_b|^{k^\parallel_a\cdot k^\parallel_b+\vec k^\perp_a\cdot\vec k^\perp_b}\prod_{a=1}^{n_c}|1-w_a\bar w_a|^{\half(k_a^\parallel)^2-\half(\vec k_a^\perp)^2} \ .
\ee

Now we wish to take into account open strings, \emph{i.e.}, $n_o \ne 0$. They only couple to the parallel parts of the fields and have the Neumann boundary conditions. The contribution is thus easily accounted for:
\be\label{eq:Kncno}
 K_{n_c+n_o} = K_{n_c+0}\prod_{n_c+1\leq a<b\leq n}|w_a-w_b|^{2k^\parallel_a\cdot k^\parallel_b}\prod_{a=1}^{n_c}\prod_{b=n_c+1}^n|w_a-w_b|^{2k^\parallel_a\cdot k^\parallel_b} \ .
\ee
Notice that there is a factor of 2 relative to bulk-bulk case in the exponents, since the
two terms in (\ref{eq:corre}) add up.

Finally, we wish to include the contribution from the boundary deformation. Since the deformation only involves the field $X^0$, we
get the contribution only from the temporal direction:
\be\label{eq:vev}
 \vev{\prod_{i=1}^N e^{X'^0(t_i)}\prod_{a=1}^{n_c+n_o} e^{i k_a\cdot X'(w_a,\bar w_a)}} =  K_{n_c+n_o}\prod_{1\leq i<j\leq N}|e^{it_i}-e^{it_j}|^2\prod_{i=1}^N\prod_{a=1}^{n_c+n_o} |w_a-e^{it_i}|^{2\xi_a} \ .
\ee
Inserting \eq{vev} to \eq{Annew}, with the expression for $K_{n_c+n_o}$ \eq{Kncno} and for $K_{n_c+0}$ in \eq{Knc}, yields \eq{Anno}.

\section{Equilibrium conditions}\label{app:equil} 

We shall look for the (global) equilibrium configuration, which is the electrostatic configuration in the Coulomb gas picture.
Let us start with the boundary $n$-point amplitude $A_{0+n}$ and set all spatial momenta are zero, $\vek_a=0$.
Since the external charges lie on the unit
circle, $\bar w_a = w_a^{-1}$. Therefore, we may
extend the Coulomb gas partition function to an analytic function of $w_a$:
\bea \label{eq:Zan}
 Z(N) &=&  \frac{1}{N!}\int\prod_{i=1}^N\frac{dt_i}{2\pi}\prod_{1\leq i<j\leq N}|e^{it_i}-e^{it_j}|^2 \nn\\
&&\times \exp\Big\{ \sum_{i,a}\xi_a\left[ \log(e^{it_i}-w_a)+ \log(e^{-it_i}-w_a^{-1})\right]\nn\\
&&+\sum_{a<b}\xi_a\xi_b\left[ \log(w_a-w_b)+ \log(w_a^{-1}-w_b^{-1})\right]\Big\} \ . 
\eea
We find that the saddle point equations can be written as
\bea \label{eq:AMeq}
 0=\inv{2}\partial_{w_c} \log Z(N) = \xi_c\left[  \sum_{a\ne c}\frac{\xi_a}{w_a-w_c} + N \left\langle\frac{1}{e^{it_1}-w_c}\right\rangle + \frac{N+\sum_{a\ne c} \xi_a}{2 w_c}\right] \ ,
\eea
where the expectation value is defined by $Z(N)$.
Notice that \eq{AMeq} is basically (the expectation value of the conjugate of) the electric force felt by the particle at $w=w_c$. The first
term is due to the self-interactions of the external charges at $w=w_a$, and the second term is the expectation value of force due to the unit
charges created by the tachyon profile. This suggests that the solution of the equations for all $w_c$ is the equilibrium configuration.

The last term in \eq{AMeq} is an electric force due to a special charge at $w=0$. The origin of this term is understood as follows.
Note that \eq{Zan} is real (for real $\xi_a$) when all $w_a$ are on the unit circle. Hence the complex derivative with respect to
any $w_c$ must return an tangential force, \ie, $\propto i w_c^{-1}$ where the proportionality constant is real. The radial force equation is
automatically satisfied when all $|w_a|=1$. In Eq. \eq{AMeq}, this is explicitly realized by the additional charge at the origin, which
cancels the radial pressure due to the interactions of the charged particles.

We may verify these arguments explicitly by splitting \eq{AMeq} into radial and tangential components.
We write
\bea \label{eq:radtan}
&&\frac{w_c}{2}\partial_{w_c} \log Z(N)+\frac{\bar w_c}{2}\overline{\partial_{w_c} \log Z(N)}  \nn\\
&=&  \inv{N}\sum_{a\ne c}\left[\frac{\xi_a w_c}{w_a-w_c} +\frac{\xi_a \bar w_c}{\bar w_a-\bar w_c}\right] + \left\langle\frac{w_c}{e^{it_1}-w_c}+\frac{\bar w_c}{e^{-it_1}-\bar w_c}\right\rangle + \frac{N+\sum_{a\ne c} \xi_a}{N} \ ; \nn\\
&&\frac{i w_c}{2}\partial_{w_c} \log Z(N)-\frac{i \bar w_c}{2}\overline{\partial_{w_c} \log Z(N)}  \nn\\
&=&\frac{i}{N}\sum_{a\ne c}\left[\frac{\xi_a w_c}{w_a-w_c} -\frac{\xi_a \bar w_c}{\bar w_a-\bar w_c}\right] + i\left\langle\frac{w_c}{e^{it_1}-w_c}-\frac{\bar w_c}{e^{-it_1}-\bar w_c}\right\rangle \ ,
\eea
where the former expression is the radial force and the latter one is tangential.
Since both $w_a$ and $w_c$ lie on the unit circle,
\beq
 \frac{w_c}{w_a-w_c} +\frac{\bar w_c}{\bar w_a-\bar w_c} = -1 \ , \ \ |w_a| = |w_c| = 1 \ ,
\eeq
\ie, the radial electric field at $w_c$ due to a particle at $w_a$ is independent of both $w_a$ and $w_c$. Hence one sees immediately that the radial component in \eq{radtan} vanishes identically, so the equilibrium configuration is fixed by the tangential equation.
If one uses rotational symmetry to fix $w_c=1$ the tangential equation becomes
\bea \label{eq:eqeq}
\inv{N}\sum_{a\ne c}\xi_a \im \frac{1}{w_a-1} + \im \left\langle\frac{1}{e^{it_1}-1}\right\rangle &=&0 \ .
\eea

Above derivation was done for fixed $N$. When the partition function is summed over $N$ we expect that the final equilibrium
equations are found by continuing analytically to $N = -\sum_a\xi_a$, in analogue with the partition function. In the text we shall
apply the equations only to such cases where the saddle point configuration is independent of $N$.

The case of nonzero $\vek_a$ is also interesting. As is easy to see from \eq{Anno}, for the boundary amplitude this means replacing $\xi_a$ by an effective charge $\xi_a^{(c)} = \xi_a + \vek_a^\parallel \cdot \vek_c^\parallel/\xi_c$ in the above formulas. Extension to bulk charges is simple as well. In this case $w_a$ and $\bar w_a$ can be taken to be independent, and the electric force obtained by differentiation has two nontrivial components.

\end{document}